\documentclass[conference]{IEEEtran}
\newcommand\blfootnote[1]{%
  \begingroup
  \renewcommand\thefootnote{}\footnote{#1}%
  \addtocounter{footnote}{-1}%
  \endgroup
}

\usepackage{amsmath,amssymb,amsfonts}
\usepackage{textcomp}
\usepackage{graphicx}
\usepackage{xcolor}
\usepackage{epsfig}
\usepackage{epstopdf}
\usepackage{algorithmic}
\usepackage{booktabs}
\usepackage{multirow}
\usepackage{rotating}
\usepackage{tcolorbox}
\usepackage{comment}
\usepackage{url}
\usepackage{cite}

\usepackage[bookmarks=false, hidelinks, breaklinks]{hyperref}

\begin{document}

\title{Surviving the Edge: Federated Learning under Networking and Resource Constraints}

\author{
\IEEEauthorblockN{Mike Mwanje, Okemawo Obadofin, Theophilus Benson, Joao Barros} 
\IEEEauthorblockA{College of Engineering, Carnegie Mellon University Africa, Kigali, Rwanda \\
Emails: \{mmwanje, oobadofi, theophib, jbarros\}@andrew.cmu.edu.}
}  

\maketitle

\blfootnote{This publication was developed as part of a program managed by Carnegie Mellon University Africa and supported by the Mastercard Foundation. The views expressed in this document are solely those of the authors and do not necessarily reflect those of Carnegie Mellon University Africa or the MasterCard Foundation.}

\begin{abstract}
Motivated by the growing proliferation of federated learning (FL) in edge environments, we present the first systematic characterization of transport-layer breaking points in FL systems operating under conditions of highly constrained network and compute resources. Using a reproducible testbed with chaos engineering tools, we evaluate Flower under progressively degraded network conditions representative of resource-constrained deployments in Africa and similar environments. Our empirical investigation reveals a fundamental mismatch between FL's burst-idle communication pattern and standard TCP connection management. We identify precise operational boundaries: FL training catastrophically fails at 5-second one-way latency due to TCP handshake timeouts, above 50\% packet loss due to buffer exhaustion, and with 90\% client dropout rates. Through systematic analysis of connection patterns during training rounds, we demonstrate that FL's periodic model update bursts, separated by extended local training periods, violate the assumptions underlying default TCP configurations. To validate the significance of these findings, we show that adjusting just three TCP connection management parameters can significantly reduce training time under extreme latency --- proving that transport-layer awareness is not merely beneficial but essential for FL deployment at the network edge. Our characterization methodology and findings provide practitioners with concrete thresholds for determining when standard FL deployments will fail and when advanced reliability techniques become necessary.
\end{abstract}

\section{Introduction}
Rapid advancements in Artificial Intelligence (AI) and Machine Learning (ML) have led to an unprecedented demand for computational resources \cite{jia2023importance}. As AI/ML continue evolving, the existing hardware struggles to meet the growing demand.

One way to meet this demand is to exploit the vast number of Internet of Things (IoT) devices that exist at the network edge. Individually, devices such as low-end smart phones, Raspberry Pis, and wireless sensor nodes, have limited potential, but their collective processing power can substantially bridge this computational gap. Moreover, the desire to train AI models on edge devices locally in Africa leads us to address unique challenges regarding connectivity. Regions such as Africa face high latencies, frequent internet shutdowns~\cite{cipesa23}, and unstable power supply~\cite{en16237708}, which together limit device availability for Federated Learning (FL)~\cite{mcmahan2017communication}. As seen in Table~\ref{tab:network-metrics}, which summarizes network latencies across continents~\cite{formoso2018deep}, Africa experiences significantly higher latencies which in some instances go even beyond 400ms. These extreme conditions are not merely outliers but represent real scenarios faced in various African regions.

\begin{table}[tb]
    \centering
    \caption{Comparison of average network latencies across continents (Adapted from \cite{formoso2018deep})}
    \label{tab:network-metrics}
        \begin{tabular}{ccl}
            \toprule
            Continent & Average Latency (ms)\\
            \midrule
            \texttt{Africa} & 280 \\
            \texttt{N. America}& 45 \\
            \texttt{Europe}& 30 \\
            \texttt{Asia}& 60 \\
            \texttt{Australia}& 50 \\
            \bottomrule
        \end{tabular}
\end{table}

While significant research has focused on algorithmic improvements to FL, the transport layer's role remains largely unexplored. Practitioners cannot make informed deployment decisions or know when advanced reliability methods are necessary without understanding where and why FL frameworks break under network stress, which is critical as FL expands into resource-constrained environments where standard assumptions often do not hold~\cite{towards-fl-at-scale}. This paper investigates the possibility of using FL effectively and efficiently when leveraging edge devices that must operate under extreme conditions. We outline our key contributions as follows:

\vspace{2pt}
\noindent
\textbf{Methodology for performance evaluation.} We designed and implemented a testbed for the rigorous evaluation of FL systems. It uses Kubernetes containers to emulate resource-constrained edge clients and integrates chaos engineering tools, to systematically inject high latency and packet loss. This facilitates the reproducible analysis of FL performance and resilience under network conditions representative of challenging real-world deployments.
    
\vspace{2pt}
\noindent
\textbf{Characterization of FL resilience.} We conducted a systematic empirical study to define the operational boundaries of a prominent FL framework, Flower~\cite{beutel2020flower}. By subjecting it to progressively extreme network conditions, including network latencies up to 10 seconds and packet loss rates up to 50\%, we demonstrate that while the framework exhibits initial robustness, performance degrades substantially under sustained network stress. Our findings quantify the tipping points beyond which effective training becomes untenable.
    

\vspace{2pt}
\noindent
\textbf{Transport-layer root cause analysis and validation.} Our investigation reveals that FL's performance degradation stems from a fundamental mismatch between its burst-idle communication pattern and standard TCP connection management, explaining why FL gRPC-based frameworks fail at specific thresholds. We validate this finding's significance by demonstrating that adjusting just three TCP parameters (\textit{tcp\_syn\_retries}, \textit{tcp\_keepalive\_time}, \textit{tcp\_keepalive\_intvl}) restores training capability where default configurations fail, confirming that transport-layer awareness is essential for FL resilience.

\section{Background and Motivation}
\label{section2}
Federated learning is an ML technique that enables training models on distributed datasets across multiple devices without exchanging the raw data. Instead of centralizing the data for training, FL allows each participating node to keep its data locally while collaboratively learning a shared model by exchanging only model updates~\cite{towards-fl-at-scale}. Although the most common industry use is to maintain data privacy, FL can be used to obtain data parallelism to aggregate several edge devices for training a model jointly, which would otherwise be impossible or take a lot of time using a single edge device.

\begin{figure}[tb]
    \centering
    \includegraphics[width=0.7\linewidth]{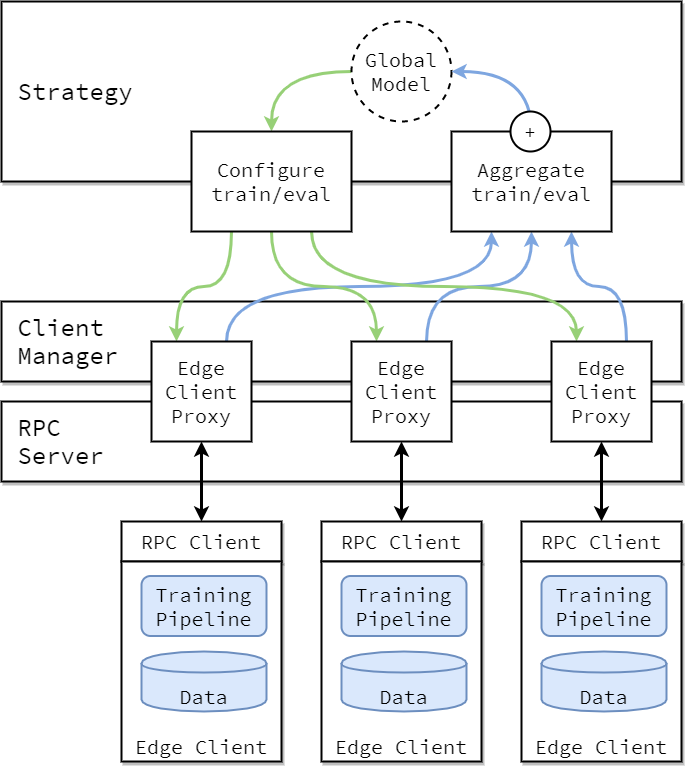}
    \caption{Architecture of the Flower Federated Learning Platform (Taken from Flower's Documentation~\cite{beutel2020flower}).}
    \label{fig:flower-architecture}
\end{figure}

We use Flower, an end-to-end FL framework that stands out due to its scalability, agnosticism, heterogeneity support, extensibility, and seamless~\cite{riedel_comparative_2024,luzon_tutorial_2024}.
Flower's  architecture (Figure \ref{fig:flower-architecture}) comprises: (1) A Server: Manages the global model, coordinates training rounds, and aggregates model updates from clients to update the global model.
(2) Clients: Perform local training on their individual data and send model updates to the server. The framework's core components include a ClientManager for handling client connections, an FL loop for coordinating the learning process, and a customizable Strategy abstraction for implementing specific FL algorithms~\cite{beutel2020flower}.

Africa faces significant challenges with network performance and reliability, often due to underdeveloped infrastructure and frequent outages. While telecom operators initially designed network technologies to tolerate maximum latencies of 300ms and 10\% packet loss in remote areas~\cite{zheleva2013increased}, much of the infrastructure is now ageing, leading to inconsistent connectivity during peak usage hours. This degradation is exacerbated by chronic power supply issues, where frequent electricity outages disrupt internet service providers and data centers, limiting consistent internet access~\cite{en16237708}. Furthermore, the region is plagued by a massive deficit in compute resources. Despite housing approximately 18\% of the global population, Africa accounts for less than 1\% of the world's total data center capacity~\cite{aiim-africa}. Finally, infrastructural limitations are worsened by political instability and interference, such as state-sanctioned internet blackouts during periods of unrest or elections~\cite{rydzak2020internet}.

These challenges manifest in poor network performance metrics across the continent, particularly in rural areas. Table~\ref{tab:network-metrics-comparison} illustrates the stark contrast between network conditions in Africa and global averages. The disparity in data centre capacity further exacerbates these network issues. For context, South Africa, leading the continent with about 100 data centres, has only 200MW of computing power, comparable to Switzerland, a much smaller country population-wise. This capacity shortage is projected to increase, with demand expected to exceed supply by 300\%, driven by growth in cloud technologies, startups, mobile financial services, and AI deployment~\cite{aiim-africa}.

\begin{table}[tb]
  \centering
  \caption{Network Metrics: Africa (Urban/Rural) vs. Global Average}
  \label{tab:network-metrics-comparison}
  \begin{tabular}{lccc}
    \toprule
    \textbf{Metric} & \multicolumn{2}{c}{\textbf{Africa}} & \textbf{Global} \\
    \cmidrule(lr){2-3} \cmidrule(lr){4-4}
    & \textbf{Urban} & \textbf{Rural} &  \\
    \midrule
    Latency (ms)        & 100--300 & 500--3000 & 50--100 \\
    Packet Loss (\%)    & 5--10    & 10--30    & $<1$    \\
    Internet Shutdowns  & Frequent & Severe    & Rare    \\
    \bottomrule
  \end{tabular}
\end{table}

In this study, we extend our experiments to test latencies as high as 10 seconds and packet loss up to 80\% to evaluate FL performance under adverse network conditions to discover where it breaks. By pushing the system to its breaking point, we can identify the specific failure modes, and lay the groundwork for developing robust optimization strategies that make FL truly practical for deployment in constrained environments.




Communication in FL is periodic --- occurring only at the end of each local training round. Assuming 10 clients, the total data transfer per round is approximately 3 MB. If updates are transmitted over 10 seconds, the required bandwidth per client is about 0.24 Mbps, with an aggregate bandwidth of 2.4 Mbps for all clients. Additionally, the asynchronous nature of FL allows clients to send updates independently, avoiding bandwidth bottlenecks that could arise with synchronous communication. It is for these reasons that our focus remains on latency, availability, and packet loss, which have the most impact on the performance of FL environments.
\section{Related Work}
\label{section3}
The most closely related works~\cite{africa-fl-for-low-resource-edge-devices,ai-deployments-in-africa} explore the difficulties of training and deploying AI models in resource-constrained environments like Africa. However, these studies typically focus on niche applications within individual countries. In contrast, our study addresses a broader infrastructural problem by systematically analyzing a wide range of network profiles observed across the continent. By evaluating this diverse spectrum of network conditions, we identify the exact failure points where advanced reliability methods provided by complementary works~\cite{towards-fl-at-scale} are required. 

Furthermore, large-scale open-source initiatives like TinyML primarily focus on optimizing AI inference on constrained hardware. Our work instead examines the systems-level challenge of \textit{training} models on limited devices over highly volatile networks. While related mobile Federated Learning (FL) approaches, such as Alibaba's Walle, address on-device training, they generally assume devices with stronger compute capabilities and relatively stable connectivity. Consequently, these existing frameworks largely overlook the severe spectrum of network degradation that our study systematically evaluates.

Many existing large-scale open source projects, e.g., TinyML, focus on optimizing AI inference within constrained devices. Our study explores an organizational problem, i.e training models on constrained devices in terrible networks. Related approaches on FL on mobile devices, e.g., Alibaba's Walle, generally focus on devices with stronger compute specs. These works often overlook the spectrum of network profiles that we examine.

There is a growing body of work on enhancing the model training aspect of FL to address a range of resource constraints from compute and network to client data and availability. These approaches provide a complementary solution to our recommendations. Similarly, work~\cite{fl-on-non-iid-data-silos} which explores the feasibility and performance of FL on non-independent and identically distributed data, provides alternative designs to address the challenges we highlight in our study. While prior studies on communication-efficient FL focus on model aggregation or compression algorithms, they generally assume a stable and transparent transport substrate. Our work differs fundamentally: we expose and quantify transport-level fragilities that arise even before model aggregation begins, highlighting the need for protocol-aware FL architectures.

\section{Characterizing FL Resilience}
\label{section4}
We begin by describing our experimental setup used for analysis, then we discuss our evaluation of Flower's behavior under ,latency and packet loss, and client failure.

\subsection{Experimental Setup}
\label{s:methodology:setup}
\begin{figure}[tb]
    \centering
    \includegraphics[width=1.0\linewidth]{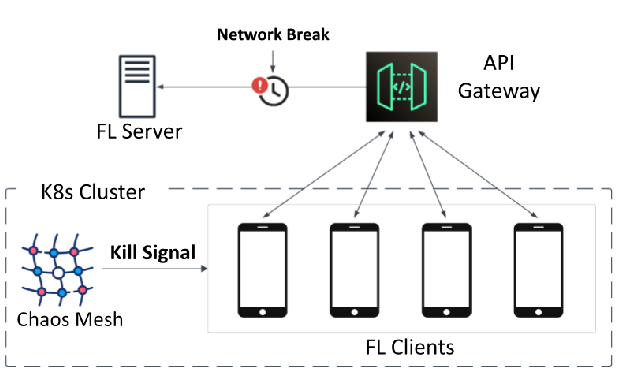}
    \caption{Federated learning test-bed}
    \label{fig:test-bed}
\end{figure}

We designed an experimental testbed that enables controlled, reproducible investigation of FL network resilience, which we used in our analysis. The platform provides a fault-injection environment comparable to large-scale network systems testbeds while maintaining full control over latency, loss, and client failure parameters. The testbed scales to tens of clients and is designed to be extensible for future FL benchmarking.
We run the FL server and FL clients on two separate networks, set up such that there was 0\% packet loss and a latency of less than 5 ms to minimize noise in our results. Given the separate networks, we could create controlled network outages. As seen in Figure~\ref{fig:test-bed}, clients run inside a containerized environment in Kubernetes, which simplifies management and collection of key metrics: network traffic, CPU, and RAM.
In configuring the clients, our goal was to allocate resources matching the lowest specification of Raspberry Pi, a 700 MHz BCM2835 model with 512 MB of RAM. To do this, we configured each client with 0.5 vCPU~\footnote{1 vCPU corresponds to 1 core running at 1 - 1.4 GHz, thus 0.5 vCPU provides a fair approximation of the Pi B’s 1-core running at 700 MHz~\cite{raspberry_pi_board_overview}.}.

To introduce network variability, we used Linux NetEm which provides control over network delays, packet loss, and bandwidth limitations~\footnote{We fixed the limit (queue size) to 200}. This involved introducing variability at the server's networking interface to enable uniform control across the various clients. Client failures were introduced using Chaos-Mesh, a chaos engineering tool for Kubernetes, enabling us to programmatically introduce client (pod) failures. To evaluate our experiments, we focus on two key metrics: \textbf{Accuracy}, which captures the model's prediction performance as reported by the Flower framework, and \textbf{Training time}, defined as the total duration required to complete a predefined number of training rounds.

\subsection{Network and Client Resilience}
By introducing high latencies, significant packet loss, and frequent client dropouts, we can identify critical thresholds where performance degrades.

\begin{figure}[tb]
    \centering
    \includegraphics[width=1.0\linewidth]{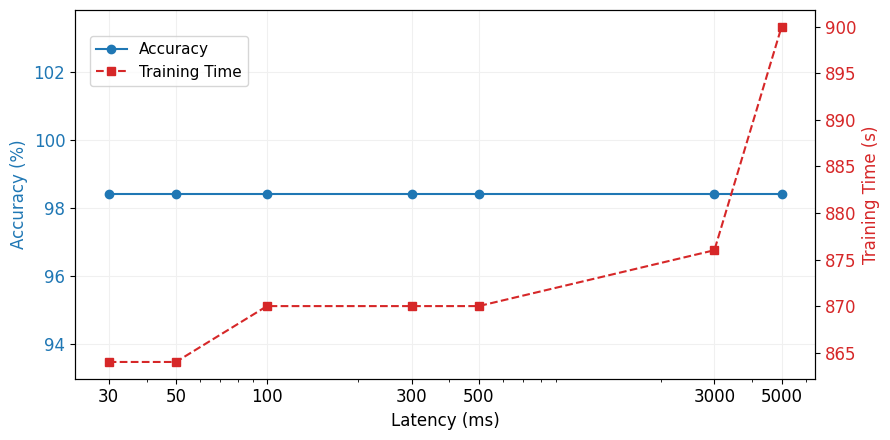}
    \caption{Impact of latency}
    \label{fig:latency-impact}
\end{figure}

\vspace{2pt}
\noindent
\textbf{Latency.} In Figure \ref{fig:latency-impact}, we present the results from varying the network latency. We observe that below 5,000ms, the key impact of increased latency is increased training time.
Latency greater than 5,000ms results in no training. In analyzing the code base we observe that this is due to a set of timeouts at the transport layer which shut down connections with significantly high latency. This is particularly problematic because Linux shuts down the connection at the start and thus prevents any communication, thus no training.

\begin{tcolorbox}[colback=gray!10, colframe=gray!80!black, boxrule=0.5pt, arc=1pt, left=2pt, right=2pt, top=2pt, bottom=2pt]
\textit{Recommendation \#1 --} Increase the default OS and framework timeout limits to tolerate extreme latency.
\end{tcolorbox}

\begin{figure}[tb]
    \centering
    \includegraphics[width=1.0\linewidth]{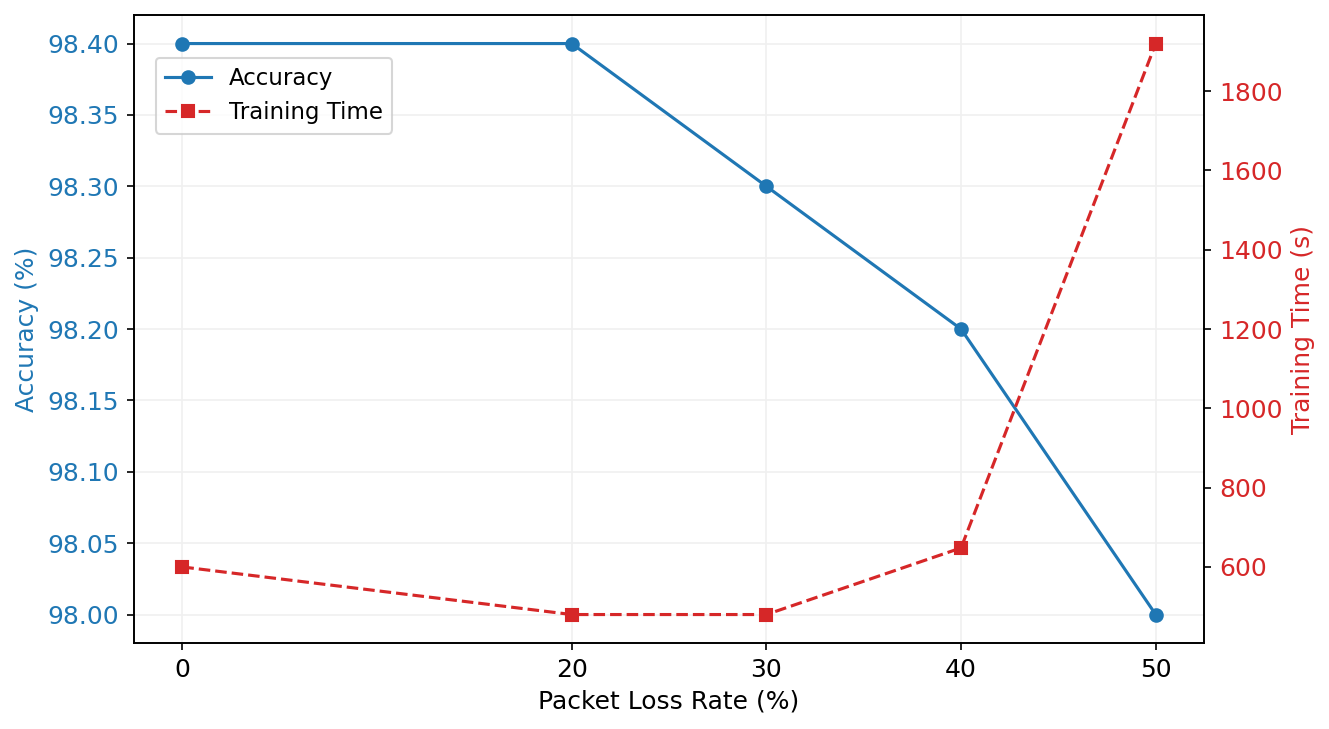}
    \caption{Impact of packet loss}
    \label{fig:packet-loss-impact}
\end{figure}

\vspace{2pt}
\noindent
\textbf{Packet loss.} From Figure \ref{fig:packet-loss-impact}, we observe three key trends: \textbf{(1)} Low packet loss (below 30\%) has no impact because the TCP contains mechanisms to re-transmit packets, effectively recovering from packet loss. \textbf{(2)} As packet loss increases from 30\% to 50\%, the model accuracy decreases by 0.3\% and training time increases by 220\%.  We observe that given the magnitude of the loss, even the retransmitted packets are lost which delays recovery. \textbf{(3)} The loss over 50\% results in the failure of Flower to train the model. Further analysis reveals that at such high loss rates, the buffers used to store packets while waiting for the lost packets do run out of space.

\begin{tcolorbox}[colback=gray!10, colframe=gray!80!black, boxrule=0.5pt, arc=1pt, left=2pt, right=2pt, top=2pt, bottom=2pt]
\textit{Recommendation \#2 --} Training failure due to packet loss can be addressed by increasing the system's buffer, although it can significantly increase model training times.
\end{tcolorbox}

\begin{figure}[tb]
    \centering
    \includegraphics[width=1\linewidth]{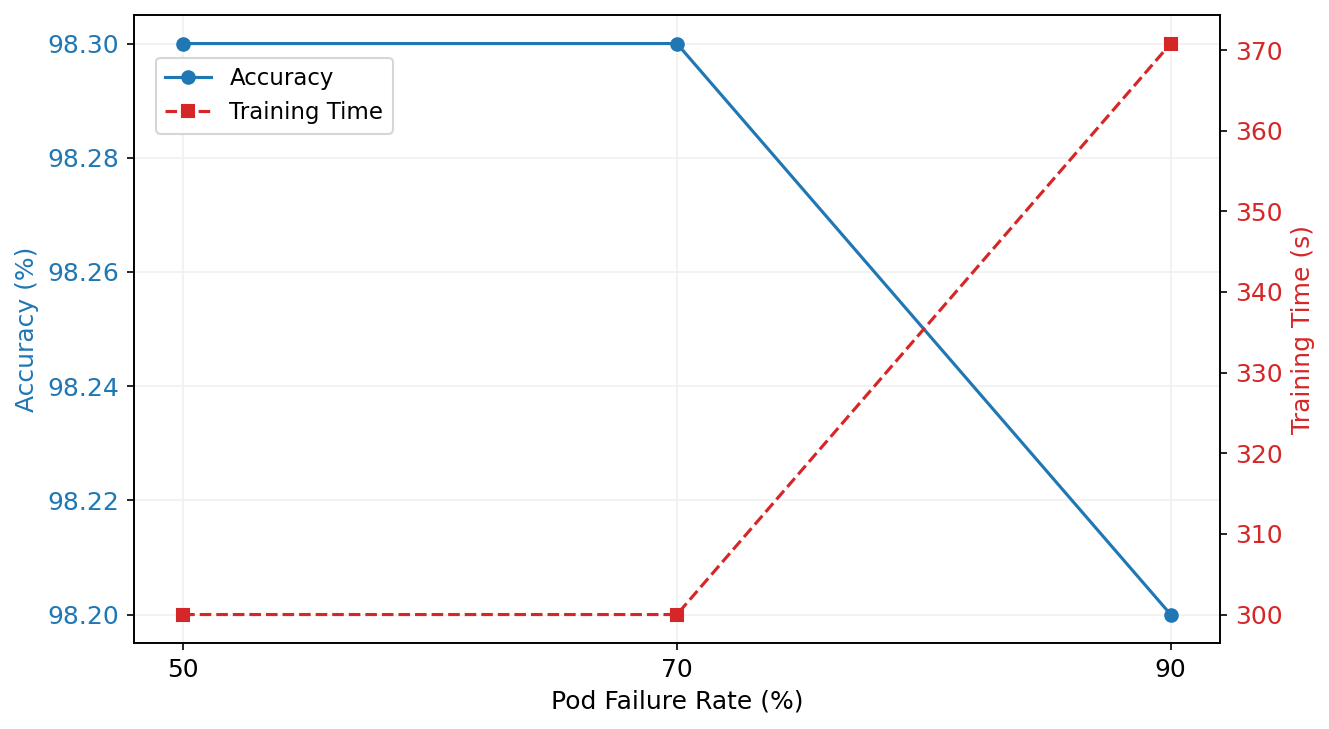}
    \caption{Impact of client failure rate}
    \label{fig:pod-failure-rate-impact}
\end{figure}

\vspace{2pt}
\noindent
\textbf{Client Failure.} Figure \ref{fig:pod-failure-rate-impact} shows the impact of pod failure on FL model training time, and accuracy. We observe no significant impact in accuracy when less than 90\% of the clients failed. Setting the minimum fit and evaluation set to 10\% allows Flower to continue training the model when at least 10\% of the clients are available thus tolerating 90\% client failure rate. However, the training time did increase by 23\% because it took longer to converge with fewer clients.

\begin{tcolorbox}[colback=gray!10, colframe=gray!80!black, boxrule=0.5pt, arc=1pt, left=2pt, right=2pt, top=2pt, bottom=2pt]
\textit{Recommendation \#3 --} Flower's resilience to client failures can be increased by lowering the minimum fit/evaluation configuration settings.
\end{tcolorbox}

Table~\ref{tab:client-fault-results} provides a summary of our analysis and characterizes regions of network performance and client outages that are acceptable, tolerable, or cause total failure. We believe that such findings can be useful to operators and developers in understanding how Flower behaves in specific environments. The tolerable region can be extended by altering Flower and/or Linux configurations to facilitate continued training in extreme conditions, the cost being longer training times.

\begin{table}[tb]
    \caption{Summary of virtual environment fault test results}
    \centering
    \begin{tabular}{cccc}
        \toprule
        \textbf{Category} & \textbf{Acceptable} & \textbf{Tolerable} & \textbf{Failure} \\
        \midrule
        Network Delay (s) & $<0.3$ & 5 & $>5$ \\
        Packet Loss (\%) & $<10$ & 30 - 40 & $>50$ \\
        Client Failure Rate (\%) & $<50$ & 50 - 70 & $>90$ \\
        \bottomrule
    \end{tabular}
    \label{tab:client-fault-results}
\end{table}

\section{TCP Parameter Tuning}
\label{section5}
Network protocol design often treats layers in isolation~\cite{4086859}. Having identified the transport-layer as the root cause of FL failures under extreme conditions, we validate our findings by testing if TCP parameter tuning can extend operational boundaries. Flower alongside many other FL frameworks rely on TCP for the reliable delivery of messages. This dependency prompted us to explore the TCP parameters available in the network configuration directory. Our goal was to identify specific settings that could validate our hypothesis regarding the benefits of cross-layer optimization. Table~\ref{s: Table 1} lists the parameters we investigated, along with a brief description of the functionality each one controls~\cite{linux_manual_tcp7_nodate}. Based on these findings, we modified our experimental testbed to include scripts that explore unique values set for each parameter, testing ranges that spanned the lower and upper bounds of the default values.

\begin{table}[tb]
    \caption{TCP Management Parameters Explored}
    \centering
    \begin{tabular}{p{3.0cm}p{5.0cm}}
        \toprule
        \textbf{TCP Parameter} & \textbf{Description} \\
        \midrule
        \texttt{tcp\_syn\_retries} & Maximum initial SYN retransmits. \\[1pt]
        \texttt{tcp\_synack\_retries} & Maximum SYN-ACK response retransmits. \\[1pt]
        \texttt{tcp\_keepalive\_time} & Idle time before keepalive probes. \\[1pt]
        \texttt{tcp\_keepalive\_intvl} & Interval between keepalive probes. \\[1pt]
        \texttt{tcp\_retries2} & Established connection retransmission limit. \\[1pt]
        \texttt{tcp\_rmem/tcp\_wmem} & Regulates socket buffer sizes. \\[1pt]
        \texttt{tcp\_max\_syn\_backlog} & Maximum queued connection requests. \\[1pt]
        \texttt{tcp\_sack} & Enables selective packet acknowledgments. \\[1pt]
        \texttt{tcp\_window\_scaling} & Enables large TCP windows. \\
        \bottomrule
    \end{tabular}
    \label{s: Table 1}
\end{table}


\vspace{2pt}
\noindent
\textbf{Packet Loss.} Tuning retransmission and buffer parameters had a negligible effect under packet loss. This was attributed to the distinct communication characteristics of FL workloads compared to traditional high-bandwidth applications. TCP parameters are primarily designed to optimize sustained data transfers with large payloads~\cite{8664630}, whereas FL communication consists of small, bursty model updates.

\begin{figure}[tb]
    \centering
    \includegraphics[width=1\linewidth]{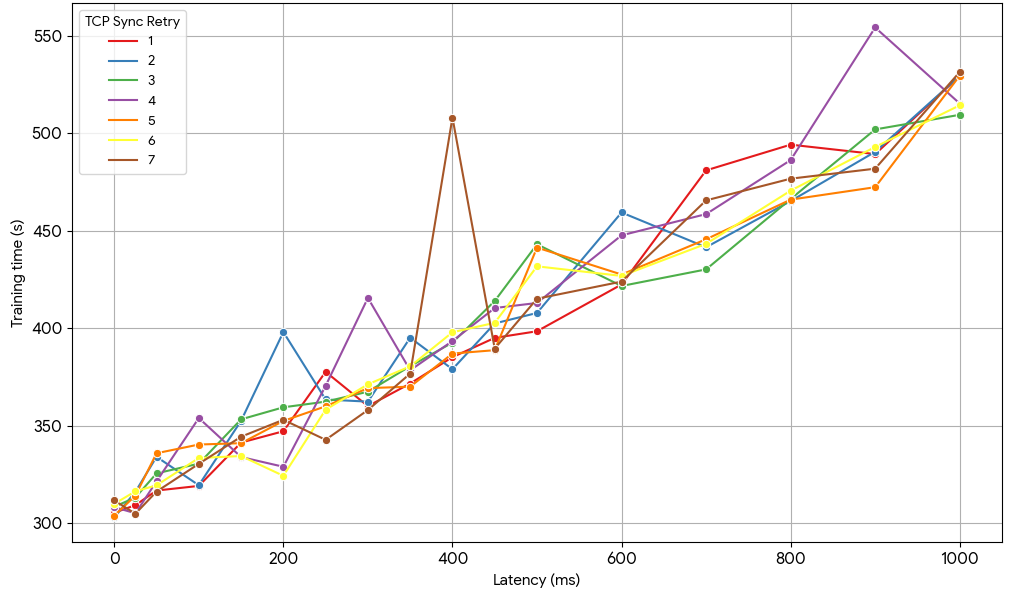}
    \caption{The impact of different \texttt{tcp\_syn\_retries} values on latency}
    \label{fig:sync_retry}
\end{figure}
\begin{figure}[tb]
    \centering
    \includegraphics[width=1\linewidth]{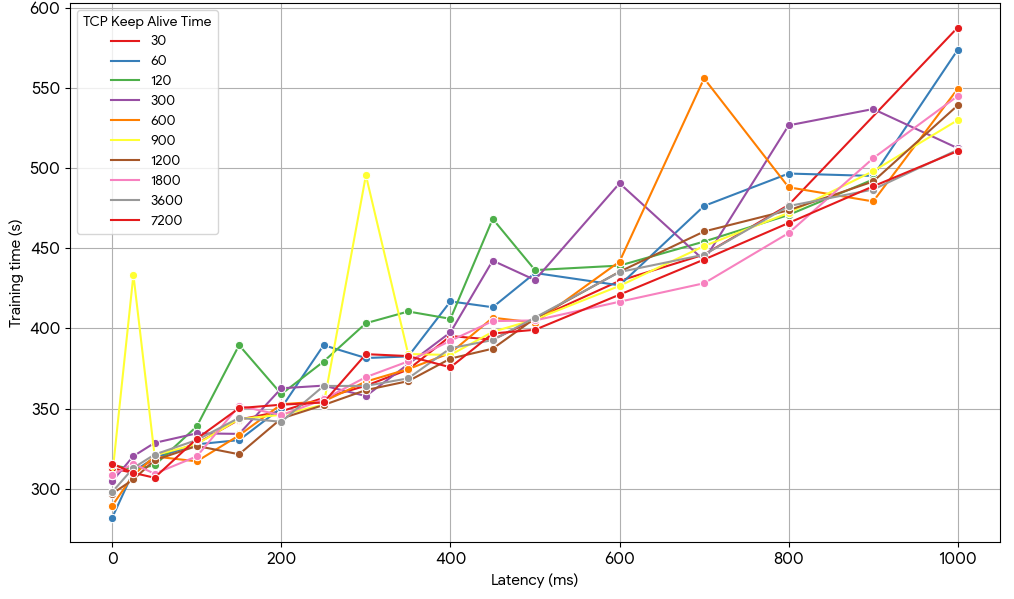}
    \caption{The impact of different \texttt{tcp\_keepalive\_time} values on latency}
    \label{fig:keep alive time}
\end{figure}
\begin{figure}[tb]
    \centering
    \includegraphics[width=1\linewidth]{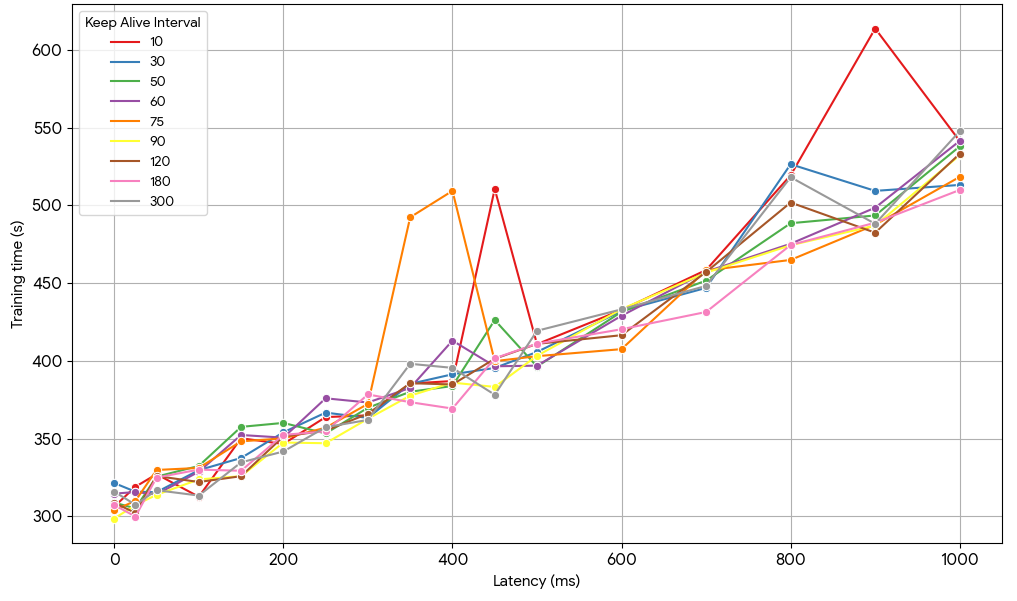}
    \caption{The impact of different \texttt{tcp\_keepalive\_intvl} values on latency}
    \label{fig:keep alive interval}
\end{figure}

\vspace{2pt}
\noindent
\textbf{Latency.} While testing latency values, we discovered three critical TCP parameters that provided substantial improvements when compared to baseline performance using their default system values. As observed in Figures ~\ref{fig:sync_retry}, ~\ref{fig:keep alive time}, and ~\ref{fig:keep alive interval} our experiments with \texttt{tcp\_syn\_retries}, \texttt{tcp\_keepalive\_time}, and \texttt{tcp\_keepalive\_intvl} revealed variable performance gains under high-latency network conditions.

In Figure~\ref{fig:sync_retry}, we can observe the impact of varying \texttt{tcp\_syn\_retries} in high-latency environments. Setting the default system value yielded suboptimal training time performance for 10 out of 17 data points (nearly 60\% of the time). For example, at latencies between 200ms and 600ms, the default \texttt{tcp\_syn\_retries} value of 6 consistently resulted in suboptimal training times. We can also observe a similar trend repeated in Figure~\ref{fig:keep alive time}. For 11 out of the 17 data points (approximately 65\% of cases), setting the default system value for the \texttt{tcp\_keepalive\_time} (7200) also resulted in longer training times. We can observe an example of this inconsistent performance at latencies between 0ms and 400ms in the figure. Finally, Figure~\ref{fig:keep alive interval} also highlights the same trend, the default setting of \textit{75 seconds} consistently fails to deliver the best overall performance for FL workloads. 12 out of 17 data points (over 70\% of our measurements) yielded reduced training times when the default \texttt{tcp\_keepalive\_intvl} values were set. Based on our evaluation, we attribute the behaviors observed in our experiments to the following underlying transport-layer dynamics:

\begin{tcolorbox}[colback=gray!10, colframe=gray!80!black, boxrule=0.5pt, arc=1pt, left=2pt, right=2pt, top=2pt, bottom=2pt]
    \textbf{Connection Establishment Resilience (tcp\_syn\_retries)} FL clients must establish reliable connections with the central server at the beginning of each training round. Under high-latency conditions (\textgreater{}100ms), the default tcp\_syn\_retries value of 6 often proves insufficient, leading to premature connection failures during the critical handshake phase.
\end{tcolorbox}

\begin{tcolorbox}[colback=gray!10, colframe=gray!80!black, boxrule=0.5pt, arc=1pt, left=2pt, right=2pt, top=2pt, bottom=2pt]
    \textbf{Connection Maintenance (tcp\_keepalive\_time/intvl)} The periodic nature of FL communication creates unique challenges for connection management. Unlike traditional web applications with continuous data streams, FL workloads exhibit bursty communication patterns during model update transmission, extended idle periods during local training phases, and sensitivity to connection drops that can invalidate entire training rounds.
\end{tcolorbox}

\section{Discussion and Future Work}
\label{section6}
Our study positions transport-layer adaptability as a first-class design axis in FL systems and other related applications of distributed systems~\cite{11050790}. These results suggest that FL frameworks should expose network-awareness hooks at runtime, allowing integration with adaptive transport modules. By treating the transport layer as part of the learning system, we shift the focus from training efficiency alone to communication resilience as a design primitive. Since prominent FL frameworks such as Flower, FATE and PySyft rely on a gRPC-based network stack, the TCP inefficiencies we identified herein likely represent a systemic issue for the broader FL ecosystem. Building on these insights, our ongoing work is exploring several promising avenues to create more resilient and efficient FL systems:

\vspace{2pt}
\noindent
\textbf{Alternative Transport Protocols:} 
We are investigating modern transport protocols, particularly QUIC, as a promising alternative to TCP for FL workloads due to its flexible congestion control algorithms. Because prominent FL frameworks like Flower rely heavily on gRPC, transitioning from TCP to QUIC could resolve many issues we observed. Furthermore, as highlighted by recent industry efforts~\cite{grpc-http3}, modern gRPC deployments are increasingly leveraging dynamic configurations to auto-adjust in response to varying latencies in real time. We plan to integrate these principles, allowing the FL transport layer to dynamically optimize its parameters based on live network conditions.

\vspace{2pt}
\noindent
\textbf{Comparative Analysis of FL Frameworks.} To understand if the observed transport-layer inefficiencies are universal, we are extending our testbed to benchmark other popular FL frameworks such as FATE and PySyft to determine whether our findings are specific to Flower's gRPC implementation or represent a more fundamental challenge for all FL systems relying on traditional network stacks.

\vspace{2pt}
\noindent
\textbf{Adaptive TCP Tuning Daemon.} We propose the design of an adaptive connection management daemon that would monitor comprehensive connection state metrics to dynamically optimize TCP parameters based on real-time network conditions.

\vspace{2pt}
\noindent
\textbf{Generalizing Across Diverse Datasets.} The efficacy of our scheme has been demonstrated on the MNIST dataset~\cite{lecun1998mnist}. To guarantee its broader applicability, we are currently evaluating our cross-layer optimizations on a wider range of datasets, including more complex image data.

\section{Conclusion}
\label{section7}
Deploying Federated Learning systems in resource-constrained environments such as those found in Africa, presents a significant challenge due to unreliable networks and limited client availability. We empirically identified transport-layer bottlenecks limiting FL resilience and showed that modest TCP tuning restores stability, motivating cross-layer co-design of future FL systems.


This work reveals a previously underexplored dependency between FL performance and transport-layer configuration, motivating the design of adaptive, cross-layer network architectures for edge learning. Our results and methodology lay the foundation for future systems that co-design learning, transport, and connectivity, enabling FL to operate reliably in the most resource-constrained environments.



\bibliographystyle{IEEEtran}
\bibliography{references}

\end{document}